\documentclass[structabstract]{aa}  
\usepackage{graphicx}
\usepackage{txfonts,pifont}
\usepackage{natbib,longtable,lscape,morefloats}
\bibpunct{(}{)}{;}{a}{}{,}
\begin{document}
   \title{On the influence of radio extended structures on offsets 
          between the optical and VLBI positions of sources in the
          ICRF2
          \thanks{Based on observations carried out with the SOAR telescope.}\fnmsep
          \thanks{Based on observations carried out with the ESO/MPG 2.2m telescope
                  during the ESO-ON agreement.}}

   \author{J.~I.~B. Camargo\inst{1,2}
           \and
           A.~H. Andrei\inst{1,2}\fnmsep
           \addtocounter{footnote}{+1}
           \thanks{Associate researcher at Observatoire de Paris/SYRTE, 77 Avenue 
                   Denfert Rochereau 75014 Paris, France.}\fnmsep
           \thanks{OATo/INAF, Osservatorio Astronomico di Torino/INAF, Strada 
                   Osservatorio 20, 10025, Pino Torinese (To), Italy.}
           \addtocounter{footnote}{-1}           
           \and
           M. Assafin\inst{2}
           \and
           \addtocounter{footnote}{+1}           
           R. Vieira-Martins\inst{1,2}\fnmsep
           \thanks{Associate researcher at Observatoire de Paris/IMCCE, 77 Avenue 
                   Denfert Rochereau 75014 Paris, France.}
           \and
           D.~N. da Silva Neto\inst{2,3}
          }

   \institute{Observat\'orio Nacional/MCT, R. Gal. Jos\'e Cristino 77,
              CEP~20921-400, RJ, Brasil\\
              \email{camargo@on.br}
         \and
              Observat\'orio do Valongo/UFRJ, Ladeira do Pedro Ant\^onio 43,
              CEP~20080-090, RJ, Brasil
         \and
              Centro Universit\'ario Estadual da Zona Oeste, Av. Manuel
              Caldeira de Alvarenga 1203, CEP~23070-200, RJ, Brasil
             }

   \date{Received ; accepted }

 
  \abstract
   {To investigate the differences between positions, as determined by optical 
    (direct imaging) and Very Long Baseline Interferometry (VLBI) techniques, 
    of extragalactic sources listed in the second realization of the International 
    Celestial Reference Frame (ICRF2).}
   {To verify the influence of the source's intrinsic structure on these
    differences.}
   {Instruments with mosaics of CCDs were used to acquire the optical positions presented
      here, leading us to opt for overlapping techniques to build a virtual, continuous CCD
      over the whole angular size of the respective fields of view, whose translation of the
      resulting intrumental positions into positions that are consistent with those in the ICRF2 was 
      made with the help of the UCAC2.}
   {The differences obtained between the optical and VLBI positions of the observed sources
      may reach more than 80 milliarcseconds and, taking into consideration that they are hardly 
      explained only by statistical  fluctuations or systematic errors in the optical reference 
      frame used here, we argue that these differences can be related to the sources' X-band
      structure index (8.4 GHz).}
   {In this context, the presence of the intrinsic structure 
   should be taken into consideration when comparing the optical and VLBI positions 
   of ICRF2 sources in the future.}


   \titlerunning{Optical positions of fundamental celestial
                 reference frame sources with radio extended structures}
   \maketitle
%

\section{Introduction}

   Fixed (non-rotating) directions on the celestial sphere can 
   be kinematically defined by the positions of very distant objects. Such a 
   concept has been the basis of the fundamental celestial frames adopted by 
   the International Astronomical Union (IAU) since 1998, starting with 
   the first realization of the International Celestial Reference Frame 
   \citep[ICRF1;][]{1998AJ....116..516M}.  The main conceptual 
   difference between the ICRF1 and its predecessor, the Fifth Fundamental 
   Catalogue \citep[FK5;][]{1988VeARI..32....1F}, is that the first had the
   directions of its coordinate axes defined kinematically by the positions
   of very distant (and therefore fixed) objects whereas fixed directions
   in the latter were given by stellar positions and proper motions based
   on the mean equator and equinox of a reference epoch -- J2000
   \citep{1998A&A...331L..33F}. 

   The current IAU's fundamental celestial frame, effective as of 1st January 
   2010, is the second realization of 
   the International Celestial Reference Frame \citep[ICRF2; for a detailed description, see][]{2009ITN....35....1F}. 
   The 
   ICRF2 contains the positions of 3414 extragalactic sources as determined by 
   Very Long Baseline Interferometry (VLBI), but only the most accurate 
   ($<$ 0.4 milliarcseconds) positions of 295 sources, selected on the 
   basis of positional stability and the lack of extensive intrinsic source 
   structure, are effectively used to define the frame axes. The remaining 
   3119 sources have positions consistent with the defining ones and help to 
   densify the frame. The ICRF2 is now the prime realization of the IAU's 
   International Celestial Reference System 
   \citep[ICRS;][]{1995A&A...303..604A}.

   In contrast with the advantage of being a natural choice to the kinematical
   definition of fixed directions, extragalactic sources may present spatially 
   extended structures at radio wavelengths. In fact, the sources' intrinsic radio 
   structure is one of the limiting factors in defining the celestial frame 
   \citep{2008IAUS..248..344C} and may also be associated to differences between the 
   orientation of the respective optical and radio frames \citep{2002AJ....124..612D}.

   The instrinsic structure of the extragalactic sources may also present temporal
   evolution. Therefore, monitoring this evolution is of 
   great importance to maintain and improve the frame. Information about the
   structure of sources in the ICRF2 can be found at the Radio Reference Frame 
   Image Database\footnote{http://rorf.usno.navy.mil/rrfid.shtml} (RRFID) and 
   the Bordeaux VLBI Image 
   Database\footnote{http://www.obs.u-bordeaux1.fr/BVID/}
   (BVID). Along with the high resolution images available therein, of 
   particular interest are the X- and S-band structure indices
   (8.4 and 2.3 GHz, respectively). This index quantifies the VLBI 
   astrometric quality of the sources with integer numbers 
   ranging from 1 (best case -- most compact sources) to 4 (worst case -- most 
   extended sources). Details about the determination of these indices are 
   found in \citet{1997ApJS..111...95F} and \citet{2000ApJS..128...17F}.
   Since the intrinsic structure may affect the alignment between the optical
   and radio frames, as mentioned above, the structure
   index is also of relevance when one thinks of the extragalactic frame
   as seen from optical wavelengths. At this point, it is just as well to
   say that this is the part of the spectrum most frequently used by
   astronomers to access the axes materialized by the ICRF2. It is
   also relevant to mention that the Gaia mission \citep{2001A&A...369..339P,2005ESASP.576....5M,2005ESASP.576...15P}
   is expected to bring the prime materialization of the celestial
   coordinate axes back to optical wavelengths and the best alignment of the future Gaia 
   optical frame with the current radio one is of great importance. For some astrometric and reference 
   frame topics related to Gaia, see \citet{2002EAS.....2..327M,
   2005ESASP.576...29L,2005jsrs.meet..196M,2008IAUS..248..217L,2009A&A...505..385A,2008A&A...490..403B,
   2010A&A...520A.113B,2011A&A...526A.102B}. For a broader variety of topics related to Gaia, see also \citet{2005ESASP.576.....T}.

   In this context, the determination of accurate optical positions of
   ICRF sources have been determined \citep[see, for instance;][]{2003AJ....125.2728A,
   2005A&A...437.1135C, 2007A&A...476..989A, 2010A&A...510A..10A}
   and provided numbers about the alignment between the optical and radio frames.
   This work focuses on statistically significant optical to radio positional 
   differences and aims at
   investigating them. 

   Here, we determine accurate optical positions of 22 sources 
   listed in the ICRF2 with the SOAR Optical Imager 
   \citep[SOI;][]{2004SPIE.5492..564S}, mounted
   at the 4.2~m aperture SOAR (SOuthern Astrophysical Research) telescope,
   and the Wide Field Imager \citep[WFI;][]{1999Msngr..95...15B} mounted 
   at the ESO/MPG 2.2 m aperture telescope. We find that statistically significant 
   optical to 
   VLBI differences in position are most frequently associated to sources with 
   large (3-4) X-band structure index. These significant differences may reach beyond 80 
   mas and are hardly explained only by systematic errors 
   affecting the astrometric reference catalogue used 
   here, the Second US Naval Observatory CCD Astrograph 
   Catalog \citep[UCAC2;][]{2004AJ....127.3043Z}, or by statistical 
   fluctuations. These results not only agree with the conclusion of 
   \citet{2002AJ....124..612D}, where a smaller mean difference was found from
   their statistic study of a larger sample, but also tell 
   that significant differences between the optical and VLBI positions of 
   reference frame sources, associated to the presence of spatially extended 
   structure, may be found on an individual basis. In Section 2, 
   we present the targets and observational procedures. In Section 3, we 
   describe the data reduction. The error budget can be found in 
   Section 4. Discussion and conclusions are presented, respectively, in 
   Sections 5 and 6.
 

\section{Targets and observational procedures}

   The 22 extragalactic sources (see Table~\ref{table1}) observed in this work 
   were selected according to their X-band structure index, as provided by 
   the BVID, and to the presence of extended structure, 
   as given by high angular resolution maps at the BVID and RRFID. The structure 
   indices shown in Table~\ref{table1} were selected by considering that the dates of the VLBI experiments should be 
   the closest to those of the respective optical observations presented here (see
   also Tables~\ref{table2} and~\ref{table3}). 
   Magnitudes
   and visibility on the sky, so that they could be observed by two southern 
   telescopes, were also taken into consideration. These telescopes were the
   SOAR (Cerro Pach\'on) and the ESO/MPG 2.2 m (La Silla), both in Chile. 
   Imagers on both telescopes are mosaics of CCDs.

\begin{table*}
\caption{ICRF2 sources that were observed with WFI and SOI - Physical characteristics and astrometric results}             
\label{table1}      
\begin{center}          
\begin{tabular}{c c c c c c c c c c c c c c c c}     
\hline\hline       
        & \multicolumn{2}{c}{\hskip -6pt Struc. Ind.} & & & & & & & &
          \multicolumn{2}{c}{\hskip  3pt(Optical$-$VLBI)} & 
          \multicolumn{2}{c}{\hskip  3pt Internal} & 
          \multicolumn{2}{c}{\hskip  3pt 3$\sigma$ level} \\
 IERS ID   & X & S & Def.? & {\it gof} &   V  & Type   &    z    & $J-K_{s}$ & Imager  &
\hskip 8pt$\Delta\alpha{*}$ & \hskip 5pt$\Delta\delta$ & 
$\sigma_{\alpha}{*}$ & $\sigma_{\delta}$ &
$\sigma_{\alpha}{*}$ & $\sigma_{\delta}$ \\ 
\hline                 
0138$-$097 & 2 & 1 & Y & -   & 17.5 & B & 0.733 & 1.547 & WFI & $+$024 & $-$042 &  9 & 18 & 72  & 86  \\  
0235$+$164 & 2 & 1 & N & 3.6 & 19.0 & Q & 0.940 & 2.084 & WFI & $+$065 & $-$015 & 16 & 19 & 82  & 88  \\
0237$-$233 & 4 & 2 & N & 2.8 & 16.6 & Q & 2.225 & 1.480 & WFI & $+$062 & $-$079 & 14 & 12 & 79  & 76  \\
0405$-$123 & 3 & 2 & N & -   & 15.3 & Q & 0.574 & 1.183 & WFI & $+$006 & $-$042 & 21 & 19 & 92  & 88  \\
0430$+$052 & 4 & 3 & N & 16.4& 14.2 & G & 0.033 & 2.089 & WFI & $-$117 & $-$064 &  6 & 27 & 69  & 105 \\
0440$-$003 & 4 & 1 & N & 3.8 & 17.0 & Q & 0.844 & 1.722 & SOI & $-$058 & $-$133 &  4 &  8 & 68  & 71  \\
0506$-$612 & - & - & Y & -   & 16.9 & Q & 1.093 & 1.402 & WFI & $-$032 & $-$065 & 15 & 19 & 81  & 88  \\
0743$-$673 & - & - & N & -   & 16.4 & Q & 1.510 & 0.770 & WFI & $+$055 & $+$018 & 18 & 27 & 86  & 105 \\
0743$-$673 & - & - & N & -   & 16.4 & Q & 1.510 & 0.770 & SOI & $+$012 & $-$012 &  7 &  5 & 70  & 69  \\
0754$+$100 & 2 & 1 & N & -   & 15.7 & B & 0.266 & 1.653 & WFI & $-$040 & $+$048 & 17 & 23 & 84  & 96  \\
0754$+$100 & 2 & 1 & N & -   & 15.7 & B & 0.266 & 1.653 & SOI & $-$086 & $+$070 &  2 &  5 & 67  & 69  \\
0808$+$019 & 1 & 1 & Y & 0.6 & 17.5 & B & 1.148 & 1.670 & SOI & $-$041 & $+$040 &  4 &  4 & 68  & 68  \\ 
0829$+$046 & 2 & 2 & N & -   & 16.7 & B & 0.174 & 1.661 & SOI & $-$028 & $+$027 &  5 &  6 & 69  & 69  \\
0906$+$015 & 3 & 2 & N & 1.4 & 17.8 & Q & 1.024 & 1.628 & SOI & $-$029 & $-$050 &  3 &  7 & 68  & 70  \\
0920$-$397 & 2 & 1 & Y & 2.7 & 18.8 & Q & 0.591 & 1.756 & SOI & $-$007 & $+$077 &  4 &  3 & 68  & 68  \\
1127$-$145 & 4 & 2 & N & -   & 16.7 & Q & 1.187 & 1.682 & WFI & $-$065 & $+$042 &  7 & 13 & 70  & 78  \\
1252$+$119 & 3 & 1 & Y & 4.2 & 16.2 & Q & 0.873 & 1.132 & WFI & $-$032 & $-$073 &  5 & 17 & 69  & 85  \\
1416$+$067 & 3 & 3 & N & -   & 16.8 & Q & 1.437 & 0.942 & WFI & $-$028 & $-$003 &  5 &  5 & 69  & 69  \\
1445$-$161 & 3 & 2 & N & -   & 18.9 & Q & 2.417 & -     & WFI & $-$050 & $-$044 & 25 & 25 & 101 & 101 \\
1622$-$297 & 3 & 2 & N & 7.0 & 20.5 & Q & 0.815 & 1.829 & WFI & $-$039 & $+$051 & 23 & 20 & 96  & 90  \\
1936$-$155 & 1 & 1 & Y & 1.8 & 19.4 & Q & 1.657 & 2.261 & WFI & $+$029 & $+$004 & 28 & 24 & 107 & 98  \\
1937$-$101 & 3 & 1 & N & 2.7 & 19.0 & Q & 3.787 & 1.461 & WFI & $+$018 & $-$053 &  9 &  8 & 72  & 71  \\
2000$-$330 & 4 & 2 & N & 8.5 & 19.0 & Q & 3.783 & 0.917 & WFI & $-$045 & $-$028 &  8 &  9 & 71  & 72  \\
2344$+$092 & 3 & 2 & N & -   & 16.0 & Q & 0.677 & 1.100 & WFI & $-$011 & $+$033 & 18 & 23 & 86  & 96  \\
\hline                   
\end{tabular}
\end{center}          
Note: all angular values are in mas. The structure indices were selected by considering 
that the dates of the VLBI experiments should be the closest to those of the respective optical 
observations presented here 
(see also Tables~\ref{table2} and~\ref{table3}). Column 1: source identification; columns 
2 and 3: X- and S-band
structure indices, respectively, as obtained from the BVID; column 4: object category (defining -- Y (yes) or not 
defining -- N (no)) indicating the object's presence in the ICRF2 defining source list 
(http://hpiers.obspm.fr/icrs-pc/); column 5:  {\it goodness of fit} ({\it gof}) as given by \citet{2001A&A...375..661G};
column 6: V magnitude; column 7: object type --  BL Lac (B), Quasar (Q), and Galaxy (G); column 8: redshift; 
column 9: near infrared colour index $J-K_{s}$ as obtained from the 2MASS;
column 10: imager name; columns 11 and 12 give the differences in the sense observed minus ICRF2 positions,
to the sources whose IDs are listed in column 1; 
columns 13 and 14 give the values of $\sigma_{I}$, as described in Section 4, to the sources whose IDs are listed 
in column 1; Columns 15 and 16 give the value of $3\sigma$, as described in Section 4, to the sources whose IDs 
are listed in column 1. 
The notation $\alpha{*}$ indicates multiplication by ${\rm cos}\delta$. Magnitudes, object type, and redshifts
were mostly taken from the ``Information on radiosources" link at http address given above. The SIMBAD was 
queried in case of missing information on the redshifts. The concerned sources are: 0237$-$233, 
1127$-$145, 1445$-$161, 2000$-$330, 2344$+$092. Regarding the magnitude, it should be noticed that these objects 
are variable.
\end{table*}

\begin{table*}
\caption{ICRF2 sources that were observed with WFI and SOI - Observational information and astrometric results}             
\label{table2}      
\begin{center}          
\begin{tabular}{c c c c c c c c c c c c}     
\hline\hline       
          & \multicolumn{2}{c}{Date of observation}
          & \multicolumn{2}{c}{Red filters} & & & &
          \multicolumn{2}{c}{\hskip  3pt Ref. Stars} & & \\
 IERS ID   & Calendar & MJD & CWL & FWHM & Exposure & Az & ZD
           & $\sigma_{\alpha}{*}$ 
           & $\sigma_{\delta}$ & \#Ref. Stars & \#Obs. \\ 
\hline                 
0138$-$097 & 2007-09-08 & 54351.215 & 651.725 & 162.184 & 35 & 63 & 36 & 37 & 46 &  67 & 22 \\ 
0138$-$097 & 2007-10-13 & 54386.331 & 651.725 & 162.184 & 35 & 286& 44 & 37 & 46 &  67 & 22 \\
0235$+$164 & 2007-10-12 & 54385.311 & 651.725 & 162.184 & 35 & 332& 51 & 43 & 41 & 131 &  9 \\
0237$-$233 & 2007-09-09 & 54352.202 & 651.725 & 162.184 & 35 & 95 & 45 & 45 & 40 &  76 & 20 \\
0405$-$123 & 2007-10-09 & 54382.243 & 651.725 & 162.184 & 35 & 63 & 31 & 41 & 45 & 132 & 12 \\
0430$+$052 & 2008-01-07 & 54472.048 & 651.725 & 162.184 & 35 & 26 & 38 & 48 & 45 & 152 &  6 \\
0440$-$003 & 2008-01-31 & 54496.127 & 628.9   & 192.2   & 50 & 307& 44 & 24 & 41 &  12 & 12 \\
0506$-$612 & 2007-10-09 & 54382.350 & 651.725 & 162.184 & 35 & 177& 32 & 41 & 38 & 186 & 12 \\
0743$-$673 & 2007-10-11 & 54384.350 & 651.725 & 162.184 & 35 & 159& 45 & 43 & 46 & 380 & 15 \\
0743$-$673 & 2007-12-07 & 54441.313 & 628.9   & 192.2   & 50 & 181& 37 & 20 & 25 &   8 &  4 \\
0754$+$100 & 2007-10-10 & 54383.389 & 651.725 & 162.184 & 35 & 42 & 49 & 42 & 43 & 408 &  5 \\
0754$+$100 & 2007-12-07 & 54441.327 & 628.9   & 192.2   & 50 & 355& 40 & 47 & 39 &   9 &  8 \\
0808$+$019 & 2007-12-07 & 54441.338 & 628.9   & 192.2   & 50 & 353& 32 & 41 & 40 &  14 &  8 \\ 
0829$+$046 & 2007-12-18 & 54452.299 & 628.9   & 192.2   & 50 & 8  & 35 & 32 & 40 &  10 &  4 \\
0906$+$015 & 2007-12-18 & 54452.817 & 628.9   & 192.2   & 50 & 14 & 32 & 45 & 27 &  11 &  4 \\
0920$-$397 & 2008-01-30 & 54495.222 & 628.9   & 192.2   & 50 & 168& 10 & 36 & 41 &  22 & 20 \\
1127$-$145 & 2007-06-08 & 54259.089 & 651.725 & 162.184 & 35 & 279& 43 & 46 & 39 & 127 & 30 \\
1252$+$119 & 2007-04-10 & 54200.188 & 651.725 & 162.184 & 35 & 358& 41 & 37 & 35 &  90 & 23 \\
1416$+$067 & 2007-04-13 & 54203.297 & 651.725 & 162.184 & 35 & 325& 42 & 44 & 38 & 143 & 25 \\
1445$-$161 & 2009-05-19 & 54970.202 & 651.725 & 162.184 & 40 & 306& 20 & 50 & 40 & 151 & 20 \\
1622$-$297 & 2007-09-09 & 54352.014 & 651.725 & 162.184 & 35 & 260& 31 & 39 & 40 & 761 & 39 \\
1936$-$155 & 2009-07-27 & 55039.620 & 651.725 & 162.184 & 60 & 54 & 21 & 45 & 48 & 523 & 12 \\
1937$-$101 & 2009-07-26 & 55038.660 & 651.725 & 162.184 & 40 & 13 & 20 & 48 & 48 & 683 & 20 \\
2000$-$330 & 2009-07-27 & 55039.653 & 651.725 & 162.184 & 40 & 113& 11 & 46 & 46 & 366 & 20 \\
2344$+$092 & 2007-09-08 & 54351.184 & 651.725 & 162.184 & 35 & 22 & 41 & 38 & 37 & 118 & 12 \\
\hline                   
\end{tabular}
\end{center}          
Note: All filter values, CWL and FWHM, are in nanometers.
Columns are, respectively: source identification, date of observation in calendar format for
year-month-day, date of observation in Modified Julian Date (MJD) format,
central wavelength (CWL) of the filter used, Full-Width at Half-Maximum (FWHM)
of the filter used, exposure time in seconds, approximate target azimuth (degrees), 
approximate target zenith distance (degrees), standard deviation in right ascension
and declination in milliarcseconds of the differences between the observed and 
catalogue positions 
for the UCAC2 stars, number of reference stars, and number of observations. Transmission 
data concerning the ESO broad band red filter (\#844) used in this work can be found at 
http://www.eso.org/sci/facilities/lasilla/instruments/wfi/inst/filters/.
Transmission data concerning the SOI red filter (R Bessel) used in this work
can be found at http://www.soartelescope.org/observing/soi-filters. Azimuth convention:
North is zero, East is $+90^{\rm o}$. Source 0138$-$097 was observed in two different dates
and all concerned CCD frames were treated together for astrometry.
\end{table*}

\subsection{SOAR Optical Imager}

   The SOI is a mosaic of two 
   2k$\times$4k CCDs separated by a gap of $7.8^{\prime\prime}$ along the 
   largest dimension. This covers a total field of view of 
   $5.3^{\prime}\times5.3^{\prime}$ with a scale of 
   $0.077^{\prime\prime}$ per pixel. A linear atmospheric dispersion 
   corrector actuated to deliver images with improved seeing. The 
   filter used throughout the observations with SOI was a R Bessel
   (see Table~\ref{table2}). 

   Overlapping images were acquired in such a way to cover these gaps and, as
   a consequence, to allow for an optical materialization of the ICRF2 axes given 
   by the largest number of UCAC2 stars available over the area covered by a 
   given mosaic. To the SOI images, the overlap could be accomplished by
   a rotation of the imager, by pointing offsets, or by a combination of these two. 
   These pointing offsets were, in some cases, applied in such a way to drive the 
   telescope towards nearby reference stars. 

\subsection{Wide Field Imager}

   The WFI is a mosaic of 4$\times$2 2k$\times$4k CCDs. A gap of $14.3^{\prime\prime}$ 
   along the right ascension axis separates the two rows of 4 CCDs, 
   running along their shortest dimension. Three gaps of $22.9^{\prime\prime}$,
   perpendicular to the one mentioned above, separate the four rows of two CCDs, 
   each gap running then along the CCD longest dimension. 
   This mosaic covers a total field of view of $33^{\prime}\times33^{\prime}$ 
   with a scale of $0.238^{\prime\prime}$ per pixel. Again, a red filter 
   (ESO844) was used throughout the observations (see Table~\ref{table2}).

   As in the case of SOI, overlapping images were acquired to cover the gaps. 
   To the WFI, the overlapping procedure was the same to all acquisitions and 
   consisted of two sets of dithered images. An offset, applied between these 
   two sets,
   displaced the targets symetrically with respect to the optical axis of the 
   imager. As a consequence, the targets were imaged in two different CCDs and 
   close to the center of the mosaic.
   
\section{Data reduction}

   In a first step, bias and flatfield corrections were applied to all
   images, from both telescopes, through IRAF\footnote{IRAF is 
    distributed by the National Optical Astronomy Observatories,
    which are operated by the Association of Universities for Research
    in Astronomy, Inc., under cooperative agreement with the National
    Science Foundation.}
   \citep{1993ASPC...52..173T}. 
   In addition, a badpixel mask was applied to the WFI data only.

   In a second step, astrometric and photometric measurements to all objects
   on the SOI and WFI images were obtained with software PRAIA \citep{2006BASBr..26...189A}
   to each CCD, individually. A mask to correct for the field distortion pattern (FDP) 
   was previously applied to the WFI observations only. The mask itself, as well as
   procedures and data used to build it, are all described in \citet{2010A&A...515A..32A}. 
   That paper is also based on WFI data and the respective observation dates 
   are close to those shown here to this same imager. No FDP mask was used on
   the SOI images since no evidence was found to justify its use.

   The UCAC3 \citep{2010AJ....139.2184Z} was released during the IAU General Assembly in Rio 
   de Janeiro, Brazil, and is supposed to supersede the UCAC2. Our option for this
   latter as reference for astrometry, however, is based on the following three reasons: 
   \citet{2010AJ....139.2440R} 
   reports problems in the proper motion system of the UCAC3; the UCAC2 was released in
   mid 2003 and since then has been widely used and analysed by a large number of works
   (up to date, more than 370 citations\footnote{According to the NASA's Astrophysics Data System 
   Bibliographic Services.}); the observational data presented in Table~\ref{table1} 
   were almost all achieved and reduced before the release of the UCAC3.

\subsection{Overlapping technique}

   Final positions to the targets listed in Table~\ref{table1} were obtained
   by means of a global reduction procedure to combine all astrometric
   information from the individual CCDs associated to the observations of a 
   given object with a given telescope.
   This procedure was discussed in details by many authors 
   \citep{1960AN....285..233E,1970MNRAS.150...35G,1992A&A...253..307B,1992A&A...264..307T,
   1998A&A...333.1107T}\footnote{The paper by \citet{1960AN....285..233E} is written
   in german and only the very enlightening english abstract is accessible to the 
   author. However, in the context of the overlapping technique, this is an important
   reference and should not be omitted.} so that we will
   give only a brief description of the calculations made here. 

   The basic equation in right ascension (declination) to an individual CCD of a given 
   mosaic is as follows:

   \begin{equation}
   \label{eq1}      
      aX+bY+c=X_{R}+DX_{R}+P_{X,Y}+r_{R},
   \end{equation}
   where 
   \[
      \begin{array}{lp{0.85\linewidth}}
         X       & is the object's abscissa on a given 
                   CCD, \\
         Y       & is the object's ordinate on a given 
                   CCD, \\
         a,b     & are constants to take rotation/shear into account on a given
                   CCD, \\
         c       & is a constant offset for a given CCD, \\
         X_{R}   & is the gnomonic projection of the object along the right
                   ascension (declination). This coordinate is read from 
                   averaging the astrometric measurements given by the second 
                   step as mentioned in the beginning of Section 3, \\
         DX_{R}  & is a correction applied to all $X_{R}$, \\
         P_{X,Y} & is a polynomial to fit the residuals on a given CCD, \\
         r_{R}   & residue.
      \end{array}
   \]
\noindent   
   In Eq.~\ref{eq1}, the unknowns are $a$, $b$, $c$, $DX_{R}$, and the coefficients 
   of $P_{X,Y}$. This polynomial is an incomplete second degree one, in the sense that 
   it contains only the terms $x^{2}$, $xy$, and $y^{2}$. This choice was made after 
   several tests involving polynomials up to the third degree.

   The determination of the unknown parameters result from a Gauss-Seidel
   iteration, where unknowns $a$, $b$, and $c$ are determined by setting 
   $DX_{R}$ and the coefficients of $P_{X,Y}$ equal to zero in the first run
   of the iteration. In the second run, $DX_{R}$ is determined from the updated
   values of $a$, $b$, and $c$. In the third run, the coefficients of
   $P_{X,Y}$ are determined from updated values of  $a$, $b$, $c$, and $DX_{R}$.
   The iteration is then implemented by going back to run one again with
   the updated values of $DX_{R}$ and $P_{X,Y}$. This procedure was shown to
   be convergent \citep{1992A&A...253..307B}. It should be noticed that the term 
   $DX_{R}$ is the responsible for interchanging information through a given mosaic, 
   since this correction is obtained from averaging positional information
   from stars that may be found in different CCDs.

   The presence of the terms $c$ and $DX_{R}$ in Eq.~\ref{eq1} gives rise to a 
   rank deficiency, so that the results delivered by the iterations described
   above are just a particular solution of the problem. Strictly speaking, this 
   rank deficiency reflects an origin indetermination so that a simple 
   translation of the whole instrumental system, as obtained from the iterations
   decribed above, suffices to provide a general solution. One way to determine 
   this translation is to elect a privileged group of stars (UCAC2 ones, for 
   instance) and force the sum of their respective final (global) corrections 
   $DX_{R}$ to be equal to zero. 

   Instead of a simple translation, however, we opted for a higher order
   polynomial to serve as boundary condition to the iterative process described
   above and to consequently derive the results shown in Table~\ref{table1}. 
   This aimed at taking into account any remaining systematic trend of the residuals,
   for the reference stars, as a function of the position on the mosaic. In the case of 
   the WFI, a complete
   third degree polynomial was employed. In the case of the SOI, a complete first
   degree polynomial was employed. In both cases, the respective coefficients were 
   determined from an iterative process that eliminated, at each iteration, the reference
   star with the largest arc length between the observed and catalogue positions. This
   star was selected among those whose absolute value of the 
   ``Observed minus Calculated'' (O-C) either in right ascension or in 
   declination was greater than 100 mas.

\section{Error budget}

   The positional accuracy to each of the observed targets was determined by
   taking into account three sources of uncertainty on their astrometric measurements: 
   the FDP, the precision of the target 
   position on the detector system, and the accuracy of the astrometric reference catalogue 
   (UCAC2) as an optical extension of the ICRF. If we call
   $\sigma_{D}$ the uncertainty from the FDP, $\sigma_{I}$ the internal precision
   of the target (that is, the precision with which the position is measured within 
   the detector system), 
   and $\sigma_{E}$ the accuracy that the optical reference catalogue materializes the coordinate
   axes of the ICRF, then the final uncertainty (at 1$\sigma$ level) of the positions of the 
   sources in Table~\ref{table1} can be given by

\begin{equation}
 \label{equation2}
 \sigma=\sqrt{\sigma_{D}^{2}+\sigma_{I}^2+\sigma_{E}^2}\hskip 5pt .
\end{equation}
   These three quantities, $\sigma_{D}$, $\sigma_{I}$, and $\sigma_{E}$, are evaluated next.

\subsection{$\sigma_{D}$}

   Figure~\ref{figure1} shows the differences, in the sense O-C, in right ascension and 
   declination for the reference stars as a function of the position on both imagers.
   Such differences indicate how well positions were corrected for the FDP.
   The results shown in this figure are typically smaller than 10 mas to the SOI and 5 mas
   to the WFI so that no systematic effects larger than these values, as a function of the 
   position on a given imager, is to be expected. It is important to notice that the error 
   bars in this figure are given by the standard deviation of the mean and the fact 
   that these bars have similar sizes comes from the uniformly distributed number of used 
   reference stars over the mosaics. In the case of the WFI, this feature would not be 
   verified without an efficient FDP model. The value adopted to $\sigma_{D}$ is, to
   both imagers, 10 mas.

\subsection{$\sigma_{I}$}

   Since each source had its position measured multiple times (see Table~\ref{table2}, last column), 
   it is natural to obtain the internal precision of this position by means of Eq.~\ref{equation3}

\begin{equation}
 \label{equation3}
 \sigma_{I}=\sqrt{\frac{1}{N-1}\displaystyle\sum\limits_{j=0}^N (P_{j}-\overline{P})^2}\hskip 10pt ,
\end{equation}
   where, to a given source, $N$ is the total number of frames that contributed to the measurement
   of both its equatorial coordinates, $P_{j}$ is one of its equatorial coordinates as measured 
   from frame $j$, and $\overline{P}$ is the respective mean value of $P_{j}$, $j=1 \ldots N$. 
   The value of $\sigma_{I}$ (or internal precision) to the observed sources is given by 
   Table~\ref{table1}, columns 13 and 14, in right ascension and declination respectively.

\subsection{$\sigma_{E}$}

   The accuracy of the UCAC2 as an optical representation of the ICRF can be obtained directly
   from \citet{2010AJ....139.2184Z}. From that paper, one finds that 20 mas is a good estimate 
   of the systematic errors in the UCAC2 positions. Therefore, 20 mas is the adopted value of 
   $\sigma_{E}$ to all observed targets. 

\subsection{Quantifying significant offsets}

   As a consequence, the confidence level, as given by columns 15 and 16 in Table~\ref{table1}, is 
   
\begin{equation}
 \label{equation4}
 3\sigma=3\sqrt{10^{2}+\sigma_{I}^2+20^2}\hskip 5pt .
\end{equation}
   We will say ``significant offset" to those offsets either in right ascension or declination, as shown in 
   columns 11 and 12 of Table~\ref{table1}, whose absolute values are greater than the respective 
   $3\sigma$ level values (Eq.~\ref{equation4}) given by the last two columns of the same table.

   The uncertainties of the VLBI positions, to the sources concerned here, are smaller than 1 mas so that
   they have negligible contribution to the error budget.

\subsection{Further considerations on errors}

\begin{figure*}
    \centering{\includegraphics[width=8cm,height=8.5cm,angle=-90]{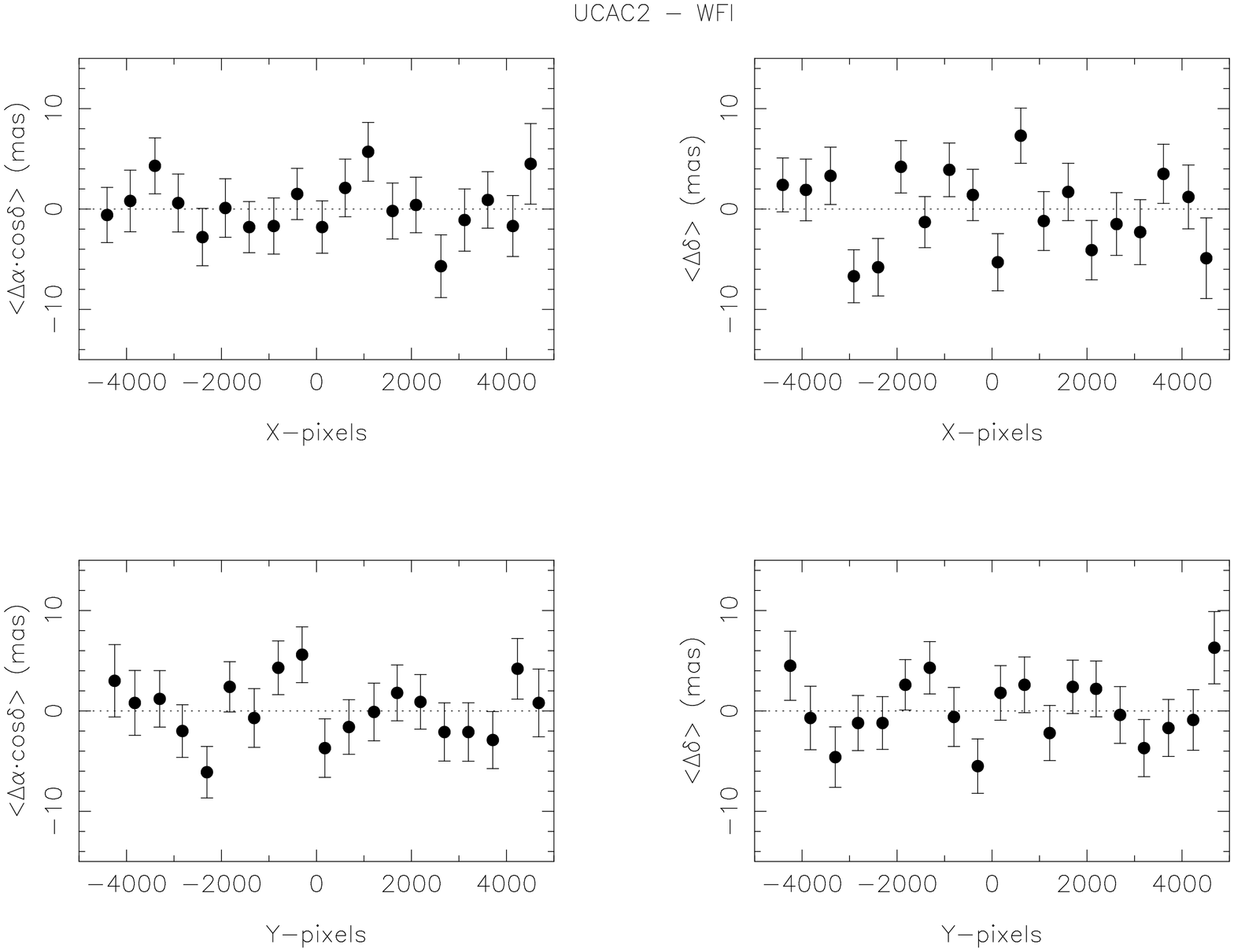}
               \hskip30pt
               \includegraphics[width=8cm,height=8.5cm,angle=-90]{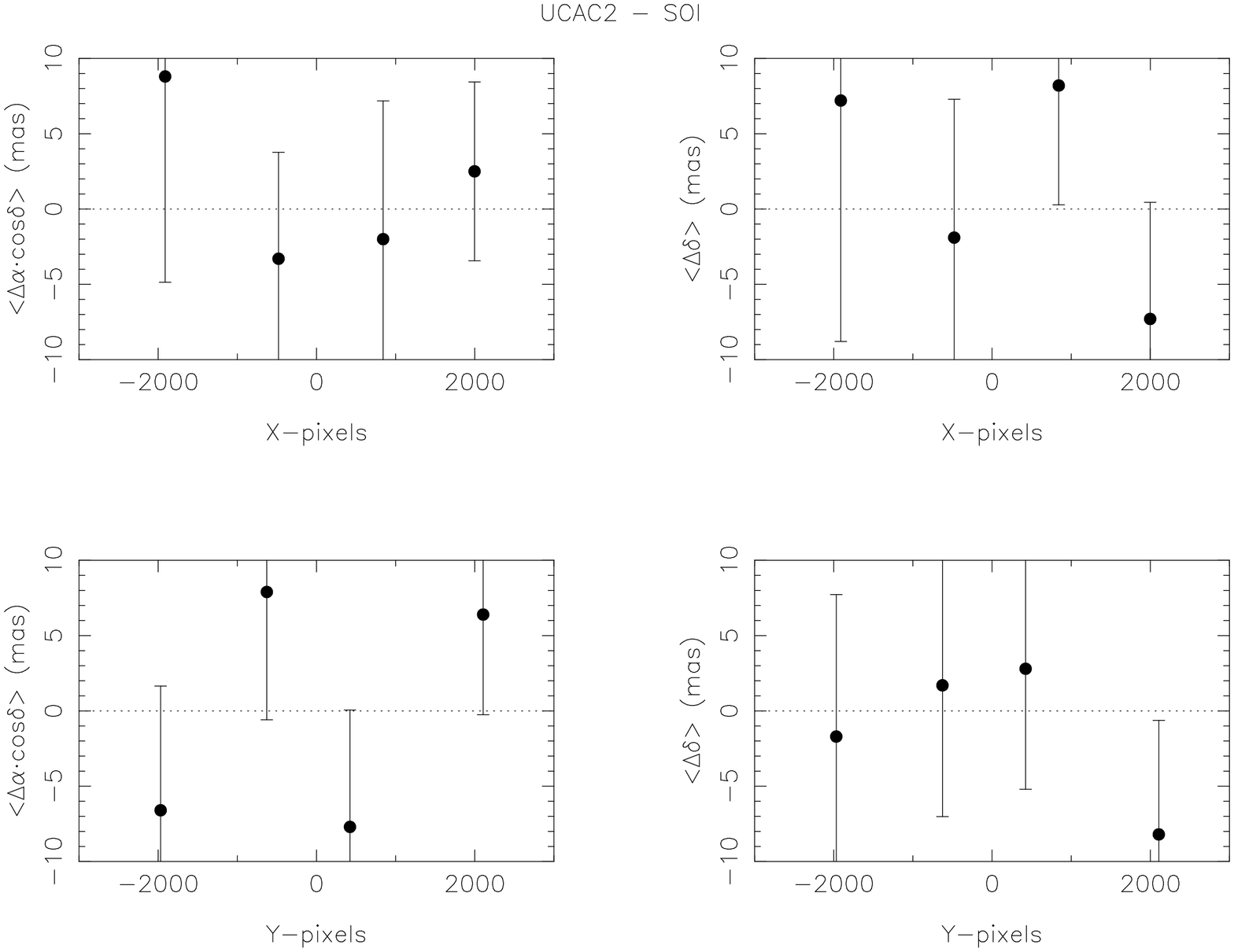}}
    \caption{Mean values of the differences in the sense O-C 
             in right ascension and declination for the reference stars, as a 
             function of the position on the whole 
             acquired field as given by the overlapping images. Bins along
             the X and Y directions encompass an arc 
             of 2.3$^{\prime}$ in the case of the WFI (left panels) and 1.8$^{\prime}$ in the case
             of SOI (right panels). Each dot represents the 
             mean of at least 100 points in the case of the WFI and 9 points in the
             case of the SOI. The error bars are standard deviations of 
             the mean and are plotted 1$\sigma$ above and below the corresponding dot.
             In these plots, the targets are always found on X$=$Y$=$0.
            }
    \label{figure1}%
   \end{figure*}

\begin{figure*}
    \centering{\includegraphics[width=7.5cm,height=8cm,angle=-90]{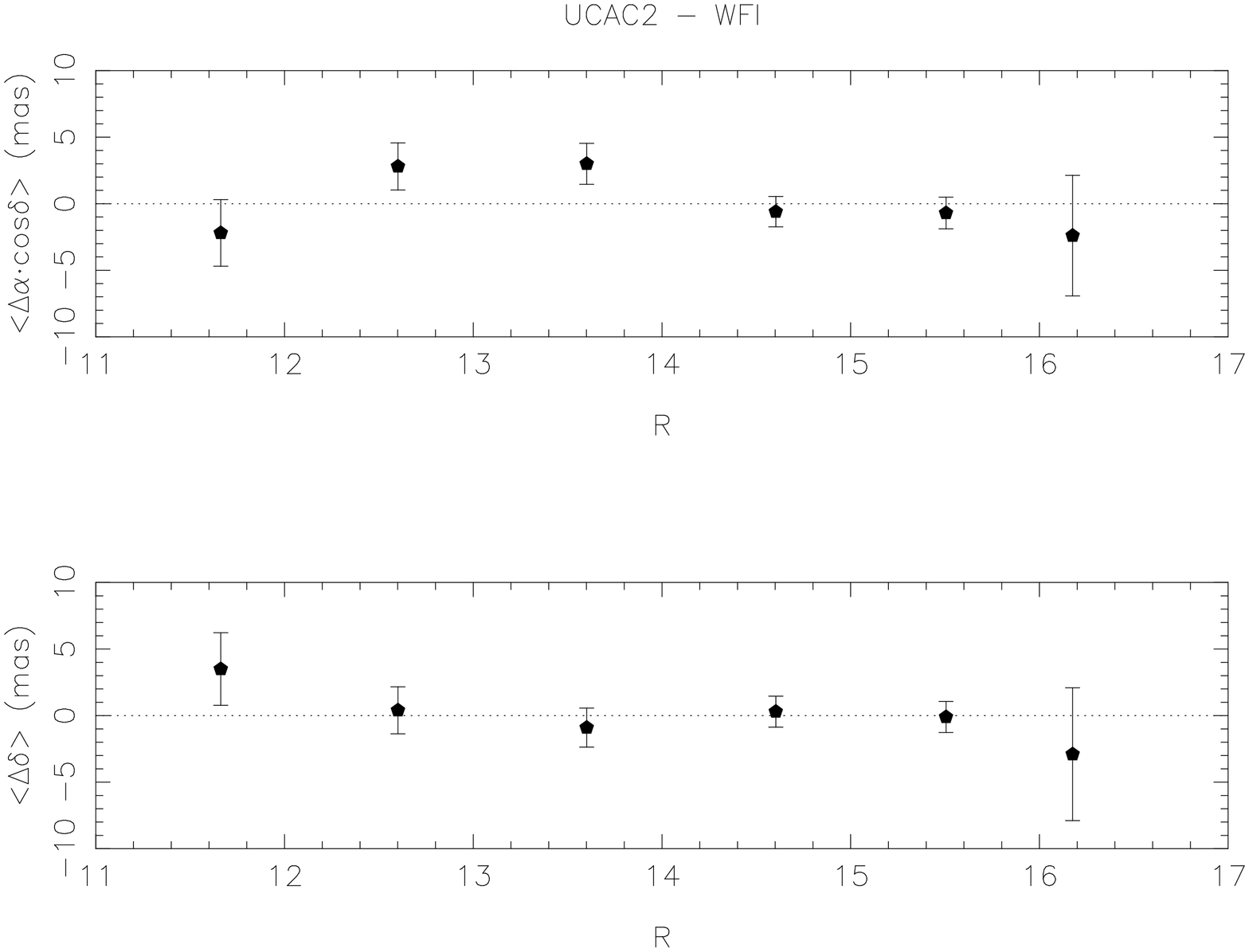}
               \hskip50pt
               \includegraphics[width=7.5cm,height=8cm,angle=-90]{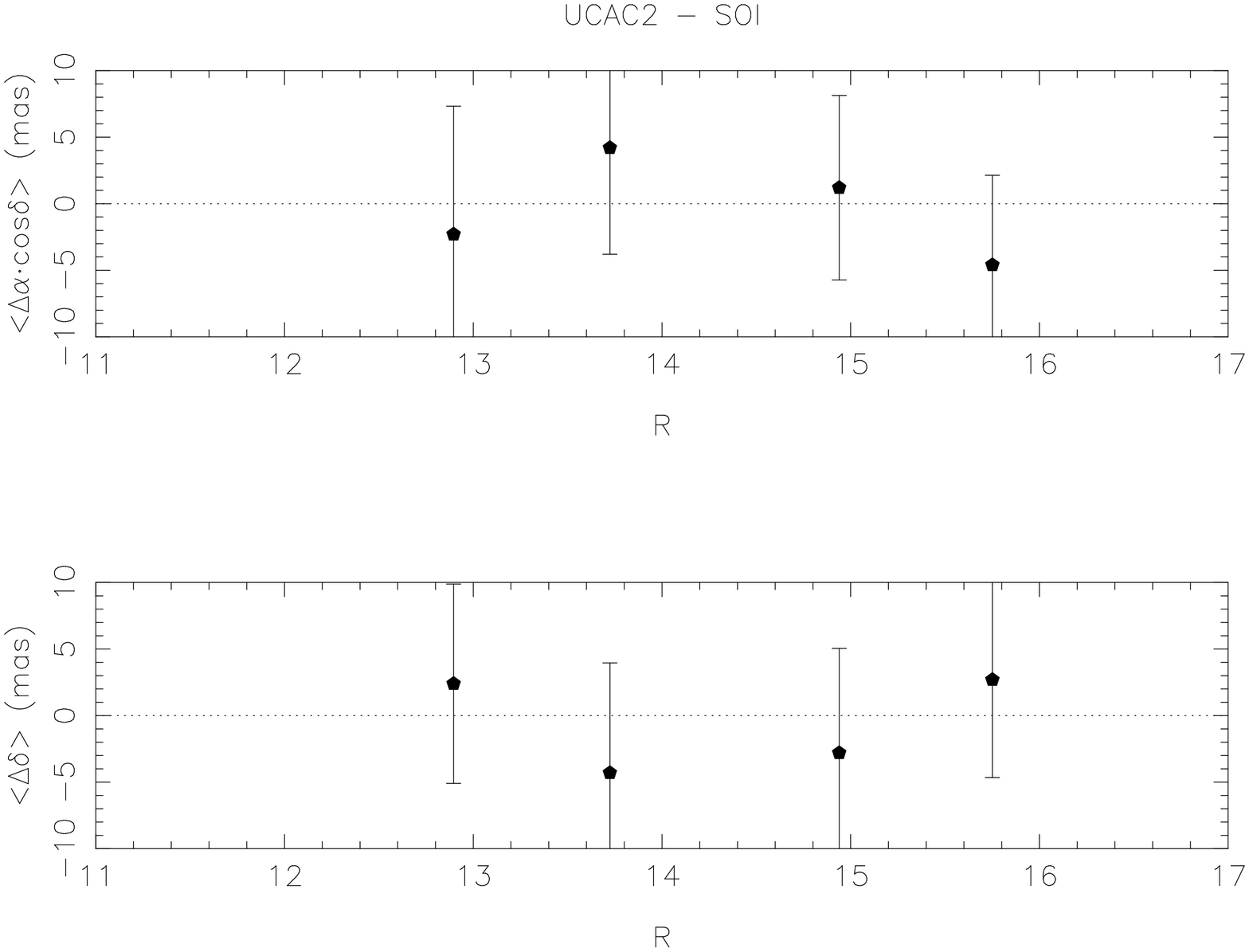}}
    \caption{Mean values of the differences in the sense O-C 
             in right ascension and declination for the reference stars. 
             Each dot represents the mean of at least 100 points in the case
             of the WFI (left panels) and 14 points in the case of the SOI (right panels). 
             The error bars are standard deviations of the mean and are plotted 1$\sigma$ 
             above and below the corresponding dot. The difference between the magnitudes of 
             the points associated to a given dot do not exceed one. The corresponding magnitude
             in the X-axis is the respective mean magnitude.}
    \label{figure2}%
   \end{figure*}

\begin{figure*}
    \centering{\includegraphics[width=7.5cm,height=8cm,angle=-90]{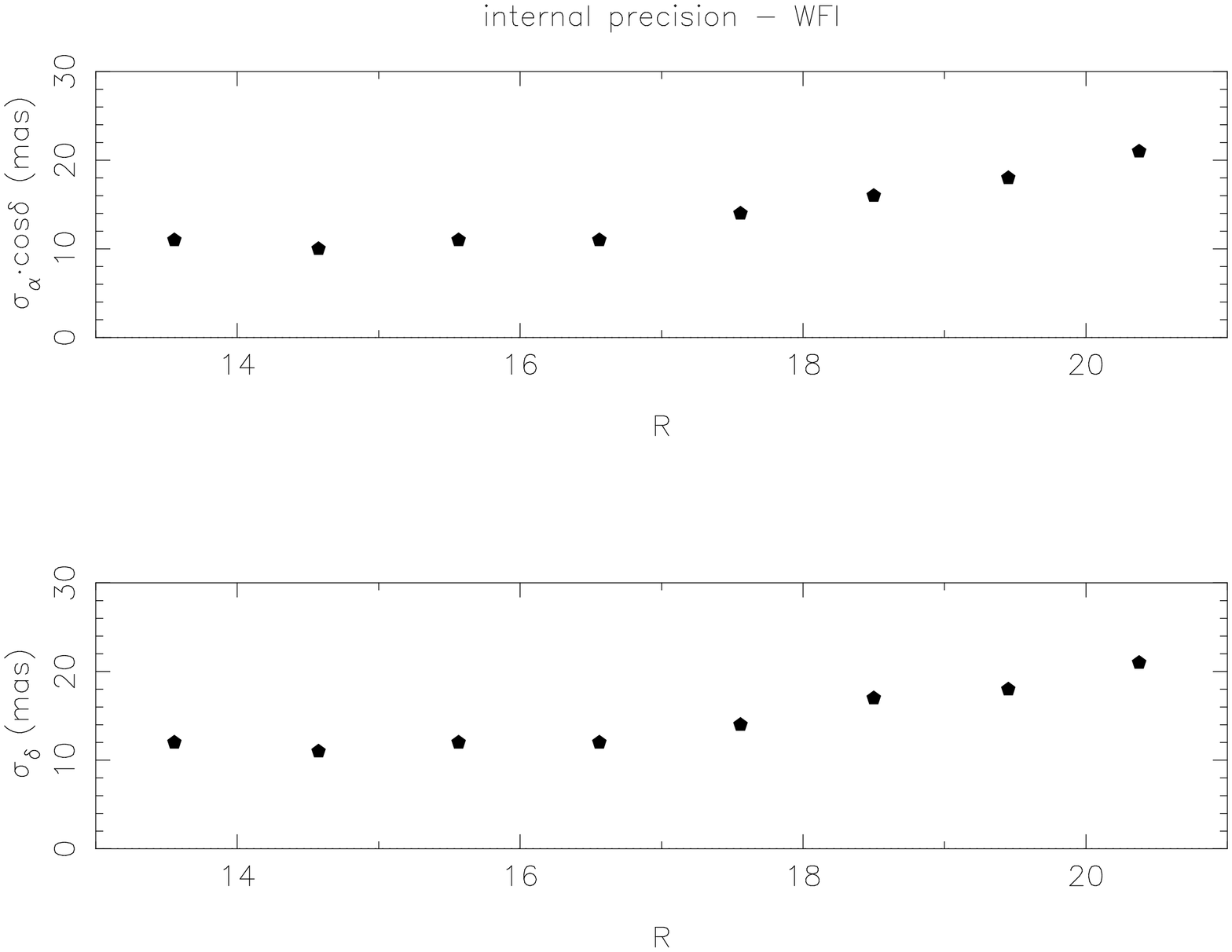}
               \hskip50pt
               \includegraphics[width=7.5cm,height=8cm,angle=-90]{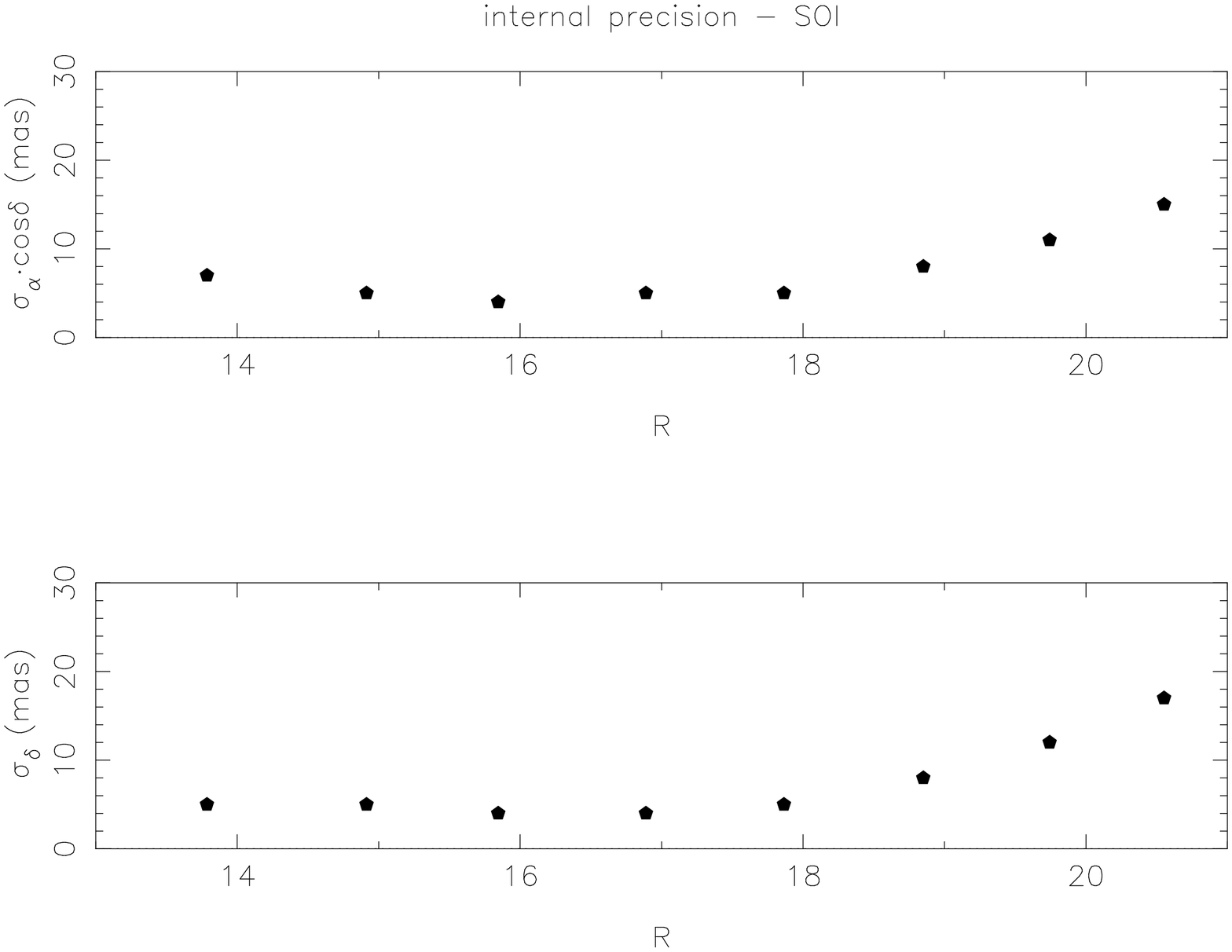}}
               \vskip40pt
    \centering{\includegraphics[width=7.5cm,height=8cm,angle=-90]{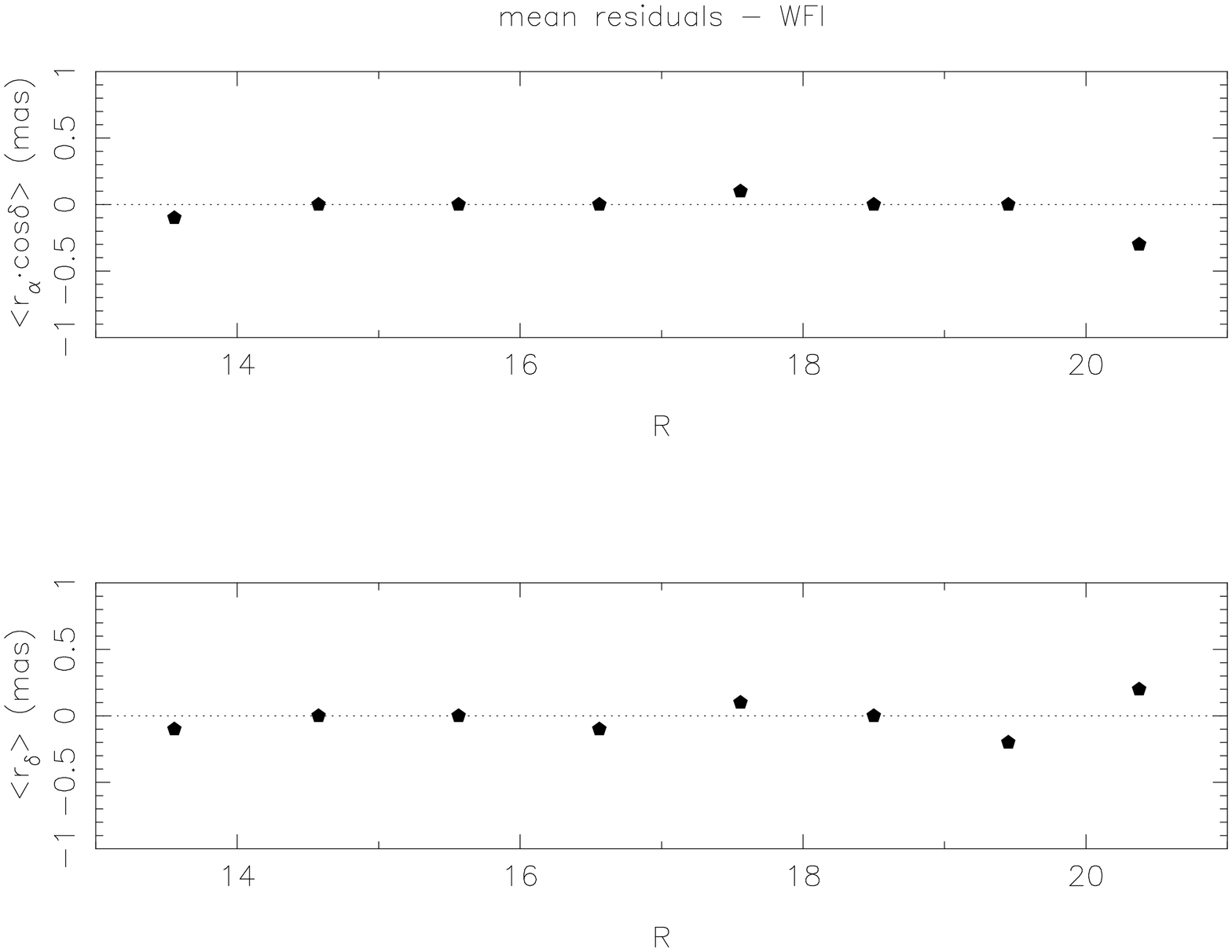}
               \hskip50pt
               \includegraphics[width=7.5cm,height=8cm,angle=-90]{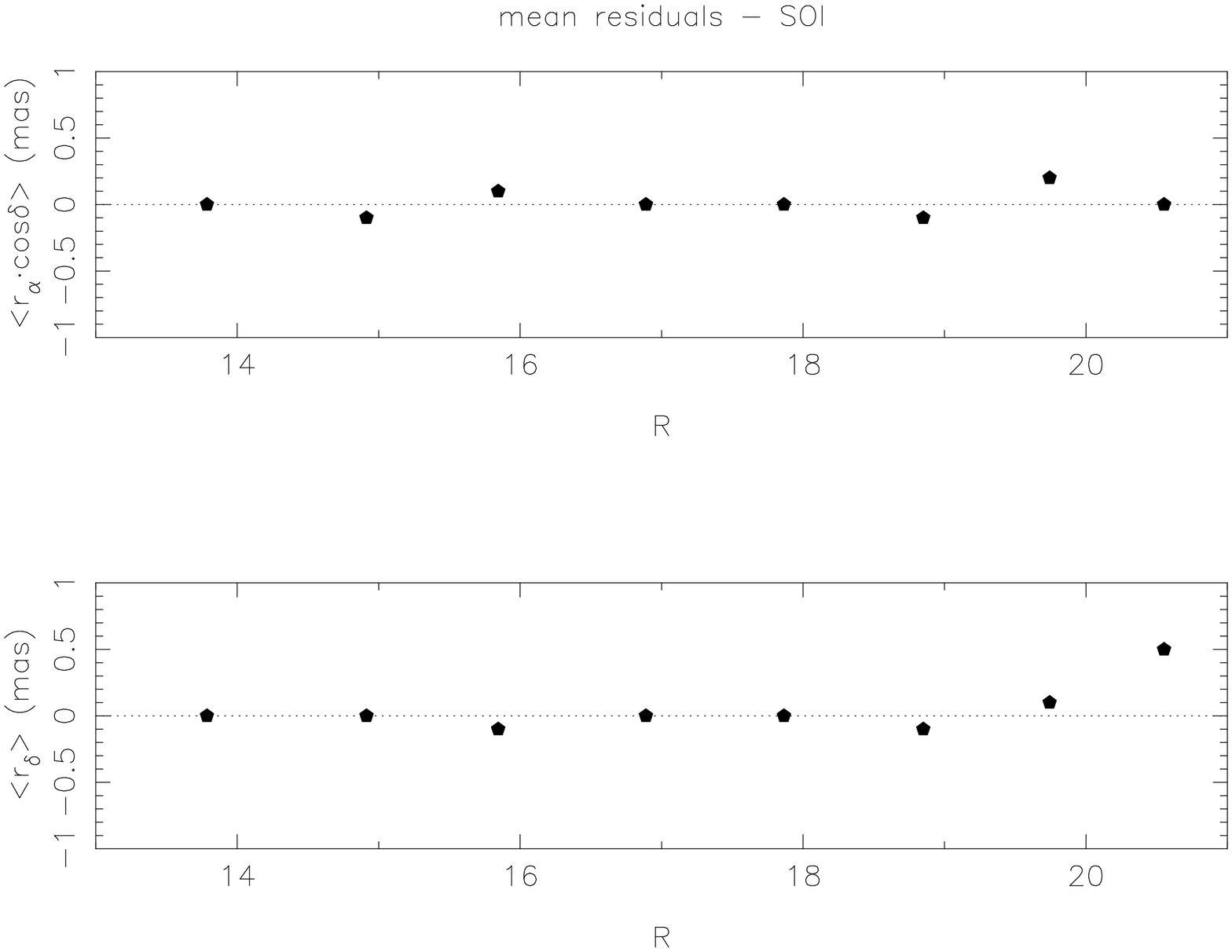}}
    \caption{Mean values of the internal precision to all objects as a function of the
             magnitude (four upper panels) and mean values of the residuals from the iterative process 
             (four lower panels), as described in Section 3, also as a function of the magnitude. 
             Each dot represents the mean of at least 13000 points in the case of the WFI
             (left panels) and 190 points in the case of the SOI (right panels). 
             The difference between the magnitudes of 
             the points associated to a given dot do not exceed one. The corresponding magnitude
             in the X-axis is the respective mean magnitude.}
    \label{figure3}%
   \end{figure*}

   The magnitudes of the reference stars used in this work are typically within the
   range $13.0<{\rm R}<16.0$. Figure~\ref{figure2} shows that systematic errors as a function
   of the magnitude are not expected to these stars.
   From Fig.~\ref{figure3} (four upper panels), one notices that the internal
   precision within this same range of magnitude is about 10 mas to the astrometric measurements 
   from the WFI and about 5 mas to those from the SOI. On the other hand, Table~\ref{table2} shows
   (columns 9 and 10) that the standard deviation of the differences between the observed and catalogue 
   positions for the UCAC2 stars is, on average, 41 mas and 39 mas to the right ascension and
   declination respectively. Taking into consideration that the observed positions of
   the reference stars present no systematic effects as a function of the magnitude, 
   and that the FDP correction was properly made (Fig.~\ref{figure1}), it is possible to conclude that
   the observed positions of the reference stars are more precise than their catalogue counterparts at
   the date of the observations carried out here.
   This is reasonable when the error propagation from the proper motions on the UCAC2 positions
   is considered.  

   In this work, the SNR was imposed by the fact that the exposure time (see Table~\ref{table2}) 
   should not saturate the UCAC2 stars as from R$\sim$13.5 to neither of the telescopes.
   This constraint is reflected in Fig.~\ref{figure3} (four upper panels), that shows the limiting best 
   internal precision achieved by each imager as well as the magnitude from which this precision 
   begins to degrade. This degradation is mostly a consequence of lower signal-to-noise ratios
   rather than effects on position due to undersampling. In fact, acquisition
   on a dither and/or offset basis of a number of images to both instruments make the
   resulting averaged (see Section 3) positions less affected by pixel phase effects
   (see, for instance,  \citet{2004AJ....127.3043Z} for a brief characterization of the 
   problem). This is strengthened by the results shown by the four lower panels of 
   Fig.~\ref{figure3}, from which no (or negligible) systematic effects on the observed positions 
   as a function of the magnitude are expected.

\begin{figure*}
    \centering{\includegraphics[width=7.5cm,height=8cm,angle=-90]{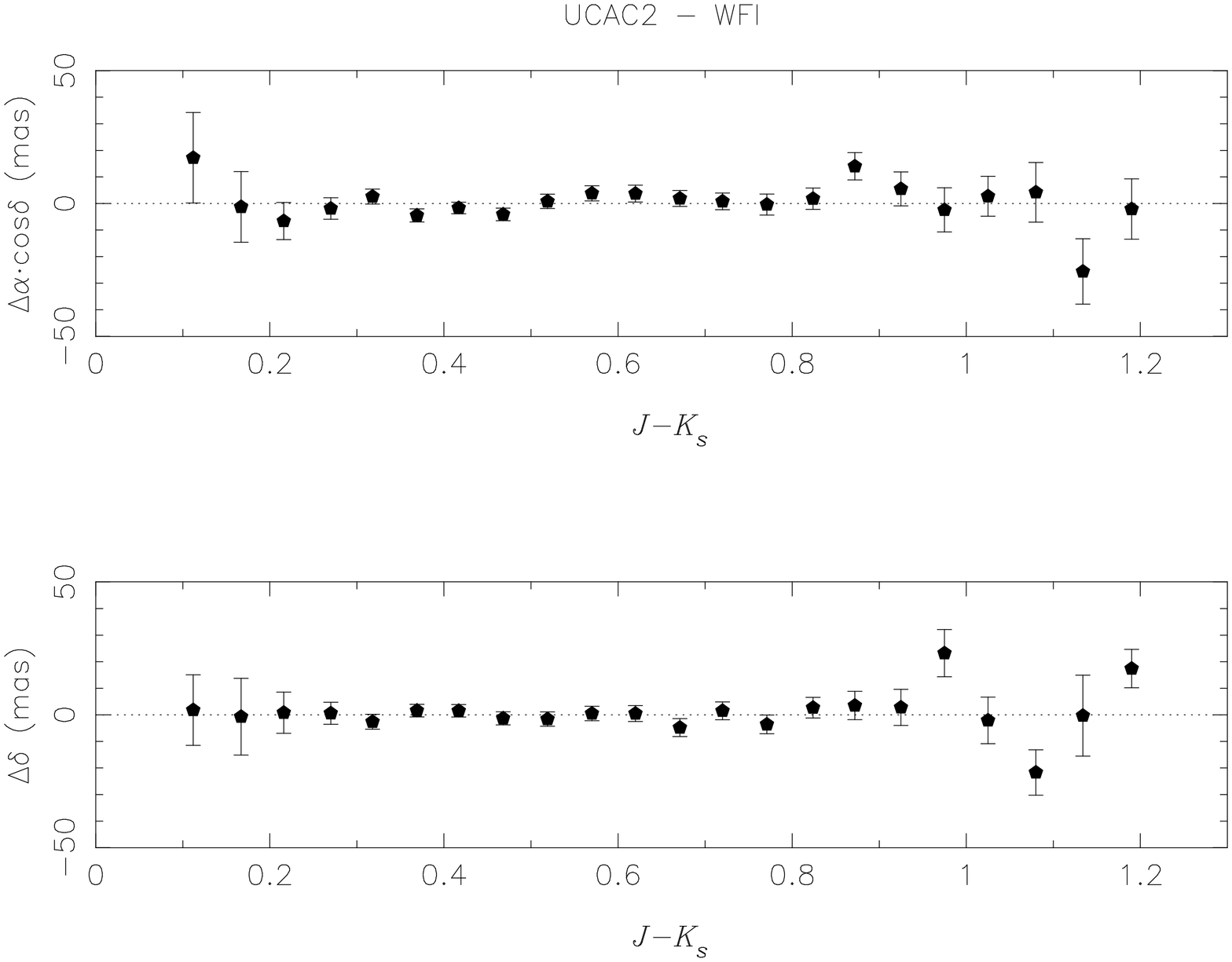}
               \hskip30pt
               \includegraphics[width=7.5cm,height=8cm,angle=-90]{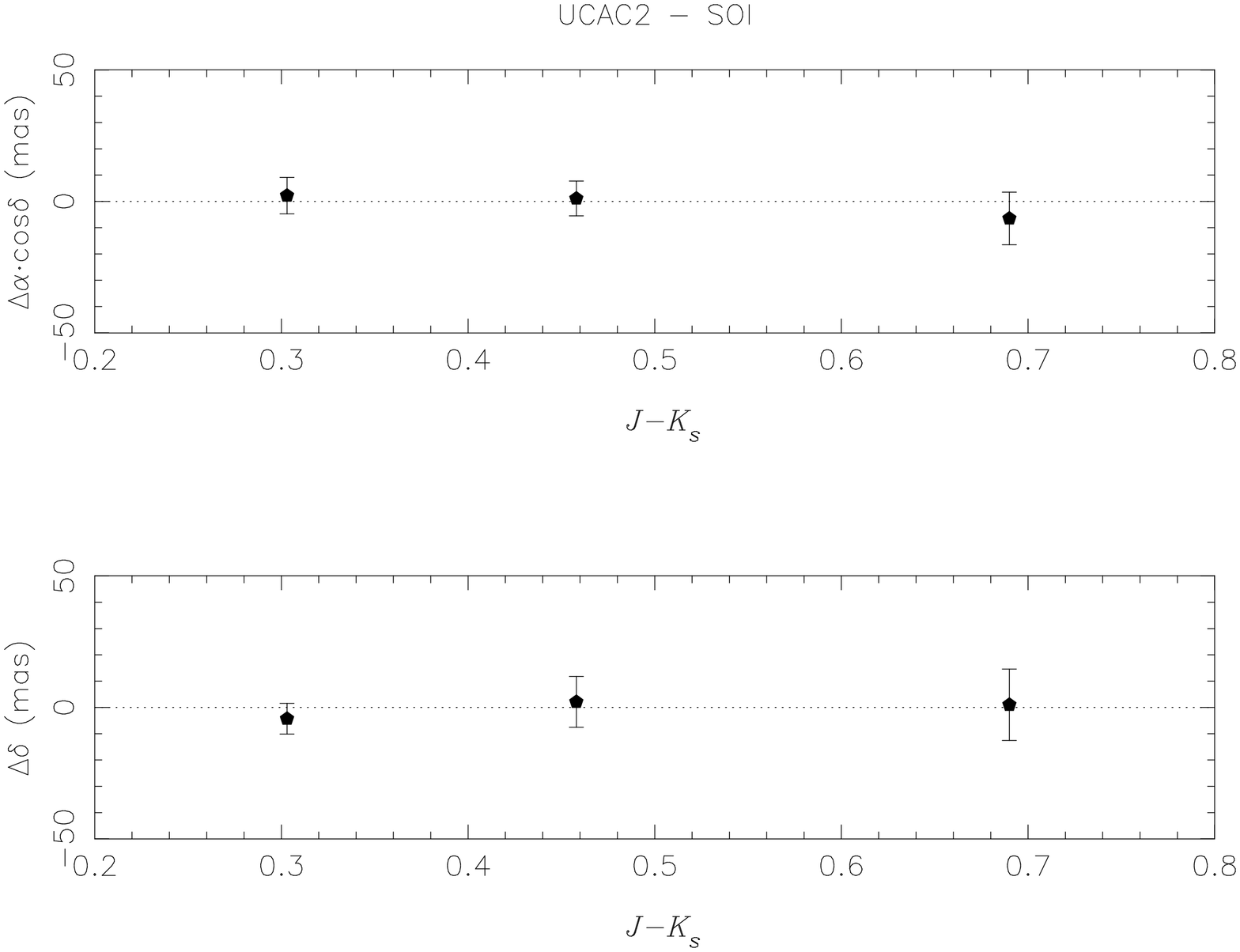}}
    \caption{Mean values of the differences in position in the sense O-C to 
             the reference stars, as a function of the $J-K_{s}$ colour index. 
             Bins along the X-axis encompass an interval of 0.05 mag  in the case of the WFI (left panels) and 
             0.2 mag in the case of SOI (right panels). Each dot represents the 
             mean of at least 12 points in the case of the WFI and 13 points in the
             case of the SOI. The error bars are standard deviations of 
             the mean and are plotted 1$\sigma$ above and below the corresponding dot.}
    \label{figure4}%
   \end{figure*}

   Another source of systematic errors on positions is related to the differential chromatic 
   refraction (DCR). Figure~\ref{figure4} tells about the DRC on the positions from the WFI
   and SOI. The results shown in this figure were derived from
   the second step described in Section 3 and point out to the fact that the final positions
   of the reference stars, as obtained from the overlapping technique, have no (or negligible)
   dependence on colour. Here, we used the  $J-K_{s}$ colour index. These near infrared 
   magnitudes are from the 2MASS \citep{2006AJ....131.1163S} and were directly obtained from the 
   UCAC2. They are considerably more precise than the visible magnitudes obtained from our observations 
   or from the observations made by UCAC2.

\begin{figure}
    \centering{\includegraphics[width=10cm,height=9cm,angle=-90]{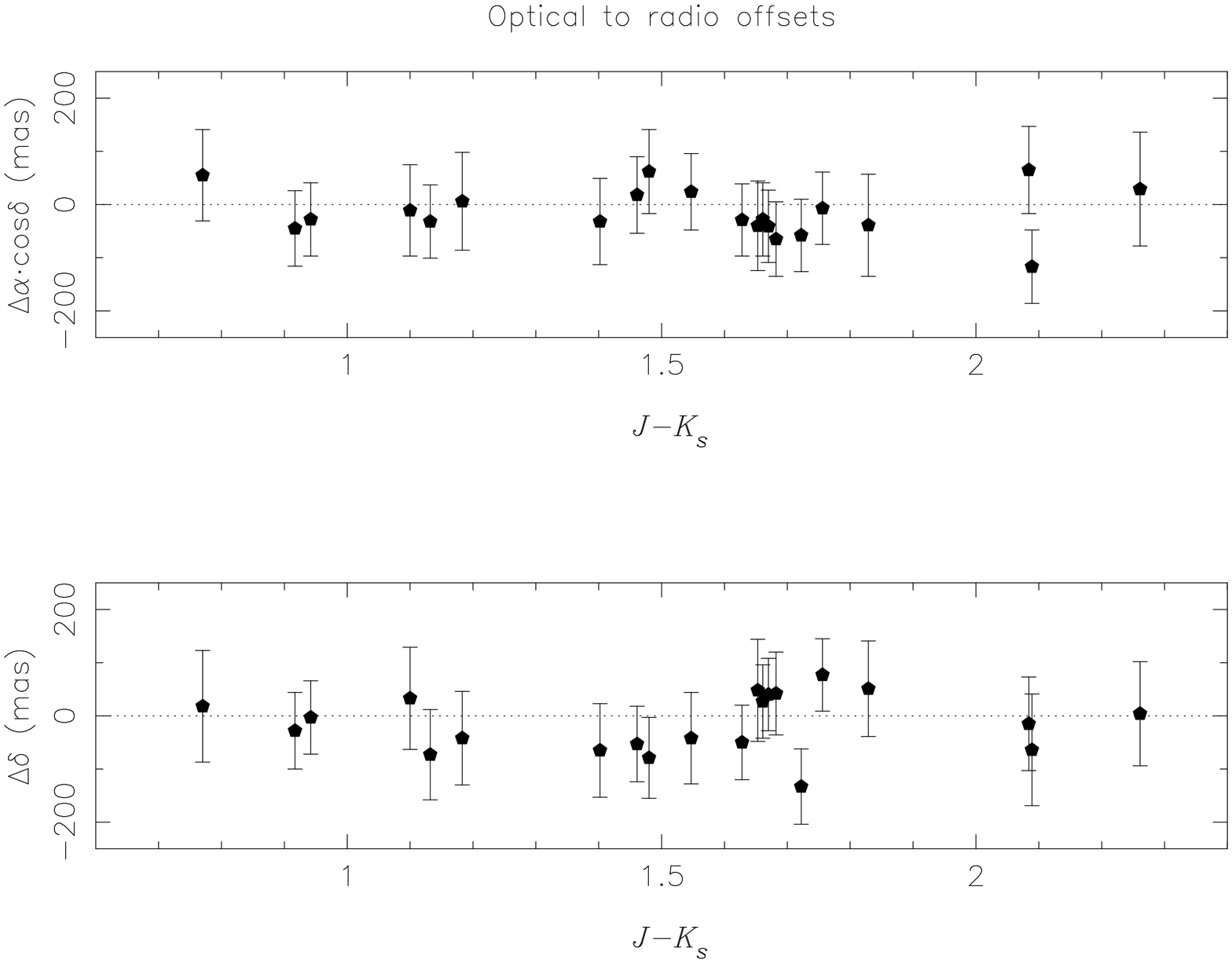}}
    \caption{Optical (observed) minus radio positions to the sources listed in Table~\ref{table1}
             (both imagers) as a function of their $J-K_{s}$ colour index 
             (column 9 of the same table). The error bars are given by the two last columns of
             Table~\ref{table1}. To the sources 0743$-$673 and 0754$+$100 only
             the results from the WFI are considered.}
    \label{figure5}%
   \end{figure}

   It should be pointed out that the reference stars are mostly main sequence objects and their average colour is 
   different from that 
   of the quasars (compare the range of colours shown in Fig.~\ref{figure4} to column 9 of Table~\ref{table1} 
   and also to the range of colours in Fig.~\ref{figure5}).
 
   Figure~\ref{figure5} plots the optical minus radio offsets in position of the observed 
   sources as a function of the respective $J-K_{s}$ colour indices. In this case, however,
   each dot in the figure represents the offset of a single source (instead of a mean offset). 
   It is important to notice that Fig.~\ref{figure5} points out no evidence of increasing
   or decreasing trend of the offset values as a function of the colour index as one would
   expect \citep[see, for instance,][]{2004AJ....127.3043Z,2006A&A...448.1235D}. This feature
   agrees with the study presented by \citet{2002PASP..114.1070S} and profited from 
   the red filters used in the observations as well as the moderate zenith distances involved. 
   In most cases, the observations were also performed not far from the local meridian 
   (see Table~\ref{table2}, columns 7 and 8).

\begin{table}
\caption{Structure index information to sources in Table~\ref{table1}, as obtained from the BVID. Dates
concern the VLBI experiments from the three most recent years.}             
\label{table3}   
\begin{center}   
\begin{tabular}{c c c c c}     
\hline\hline       
 IERS ID     & Experiment    & Date        & X & S \\ 
\hline                 
0138$-$097  & RDV67          & 2008-01-23  & 2 & 1 \\
            & RDV59          & 2006-09-13  & 2 & 1 \\
            & BL115C\_SX     & 2004-02-15  & 3 & 1 \\
\\
0235$+$164  & RDV71          & 2008-09-03  & 1 & 1 \\
            & RDV66          & 2007-12-05  & 2 & 1 \\
            & RDV65          & 2007-08-01  & 1 & 1 \\
            & RDV64          & 2007-07-10  & 2 & 1 \\
            & RDV62          & 2007-03-27  & 2 & 1 \\
            & BLL122B\_SX    & 2005-08-26  & 1 & - \\
            & RDV51          & 2005-06-29  & 2 & 1 \\
\\
0237$-$233  & RDV70          & 2008-07-09  & 4 & 2 \\
            & RDV62          & 2007-03-27  & 4 & 2 \\
            & BR945          & 1994-07-08  & 4 & 3 \\
\\
0405$-$123  & RDV62          & 2007-03-27  & 3 & 2 \\
            & RDV57          & 2006-07-11  & 3 & 2 \\
            & RDV20          & 2000-03-13  & 2 & 2 \\
\\
0430$+$052  & RDV68          & 2008-04-02  & 4 & 3 \\
            & RDV42          & 2003-12-17  & 4 & 3 \\
            & RDV31          & 2002-01-16  & 4 & 2 \\
\\
0440$-$003  & RDV7           & 1998-02-09  & 4 & 1 \\
            & VABF12         & 1995-10-12  & 1 & 1 \\
\\
0754$+$100  & RDV71          & 2008-09-03  & 2 & 1 \\
            & BL122B\_SX     & 2005-08-26  & 2 & - \\
            & BL115C\_SX     & 2004-02-15  & 2 & 1 \\
\\
0808$+$019  & RDV72          & 2008-12-17  & 2 & 2 \\
            & RDV70	     & 2008-07-09  & 2 & 2 \\
            & RDV69          & 2008-05-14  & 1 & 1 \\
            & BL122B\_SX     & 2005-08-26  & 1 & - \\
            & BL115C\_SX     & 2004-02-15  & 1 & 1 \\
\\
0829$+$046  & RDV61          & 2007-01-24  & 2 & 2 \\
            & RDV57          & 2006-07-11  & 3 & 1 \\
            & B9725A         & 1997-01-10  & 3 & 2 \\
\\
0906$+$015  & B9725A         & 1997-01-10  & 3 & 2 \\
\\
0920$-$397  & RDV72          & 2008-12-17  & 2 & 1 \\
            & RDV67          & 2008-01-23  & 2 & 1 \\
            & RDV65          & 2007-08-01  & 1 & 1 \\
            & RDV42          & 2003-12-17  & 3 & 1 \\
\\
1127$-$145  & RDV20          & 2000-03-13  & 4 & 2 \\
            & BR945          & 1994-07-08  & 4 & 2 \\
\\
1252$+$119  & RDV67          & 2008-01-23  & 2 & 1 \\
            & RDV61          & 2007-01-24  & 3 & 1 \\
            & B9725A         & 1997-01-10  & 2 & 1 \\
\\
1416$+$067  & RDV66          & 2007-12-05  & 3 & 3 \\
            & RDV8           & 1998-04-15  & 2 & 1 \\
            & BR9525         & 1995-04-12  & 2 & 1 \\
\\
1445$-$161  & RDV31          & 2002-01-16  & 3 & 2 \\
            & B9725A         & 1997-01-10  & 3 & 2 \\
\\
1622$-$297  & RDV70          & 2008-07-09  & 3 & 2 \\
\\
1936$-$155  & RDV71          & 2008-09-03  & 1 & 1 \\
            & RDV59          & 2006-09-13  & 2 & 1 \\
            & BL115C\_SX     & 2004-02-15  & 1 & 1 \\
\hline                   
\end{tabular}
\end{center}
\end{table}

\addtocounter{table}{-1}
\begin{table}
\caption{Structure index information to sources in Table~\ref{table1}, as obtained from the BVID -- cont.}             
\begin{center}   
\begin{tabular}{c c c c c}     
\hline\hline       
 IERS ID     & Experiment    & Date        & X & S \\ 
\hline                 
1937$-$101  & RDV57          & 2006-07-11  & 3 & 1 \\
            & B9725A         & 1997-01-10  & 3 & 1 \\
\\
2000$-$330  & RDV69          & 2008-05-14  & 4 & 2 \\
            & RDV62          & 2007-03-27  & 4 & 1 \\
\\
2344$+$092  & RDV31          & 2002-01-16  & 3 & 2 \\
            & B9725A         & 1997-01-10  & 3 & 2 \\
\hline                   
\end{tabular}
\end{center}
Note: The information in columns from two to five were taken from 
the BVID. Columns are: source identification, experiment, experiment date, 
and X- and S-band structure indices, respectively.
\end{table}

\section{Discussion}

\begin{figure}
    \centering{{\includegraphics[width=7cm,height=9cm,angle=-90]{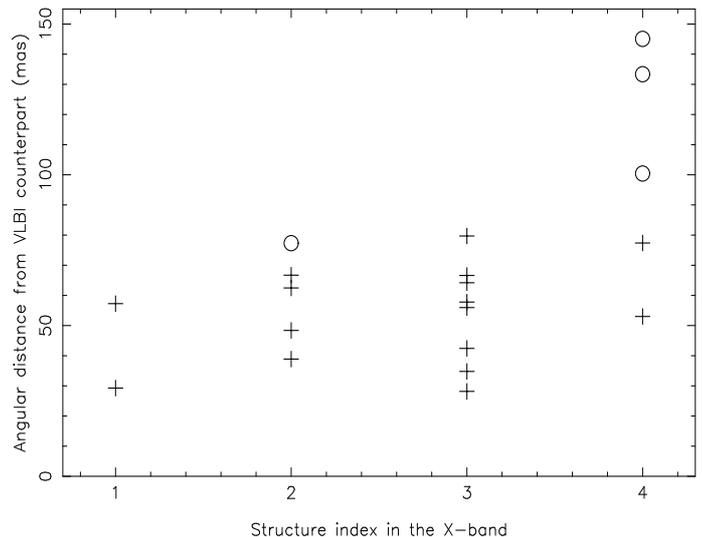}}}
    \caption{Angular distance (see Section 5) between the VLBI positions and the respective
             optical ones determined here as a function of the X-band structure index.
             Circles: sources to which at least one of the offsets, $\Delta\alpha{*}$ 
             or $\Delta\delta$ (Table~\ref{table1}, columns 11 and
             12), were significant (Table~\ref{table1}, columns 15 and 16 and Section 4).
             Crosses are given otherwise.}
    \label{figure6}%
   \end{figure}

   From Table~\ref{table1}, one notices that four sources presented significant offsets. They
   are 0237$-$233, 0430$+$052, 0440$-$003, and 0920$-$397. In Fig.~\ref{figure6}, the angular distance 
   ($\sqrt{\Delta\alpha{*}^{2}+\Delta\delta^{2}}$) between the respective optical and VLBI 
   positions of the observed sources is given as a function of the 
   X-band structure index, where $\Delta\alpha{*}$ and $\Delta\delta$ are given 
   by columns 11 and 12 of Table~\ref{table1}. These four sources are indicated with a circle
   in Fig.~\ref{figure6}.
   The other sources are indicated with a cross.

   It should be noticed in Fig.~\ref{figure6} that the structure index 4 hosts
   the majority of the sources with significant offsets, as indicated by a circle.

   We can set, from columns 15 and 16 in Table~\ref{table1}, a conservative average threshold of
   80 mas from which optical to radio offsets are considered to be significant in
   this work.
   This value is larger than the positional accuracy of 60 mas presented by 
   \citet{2010A&A...515A..32A}, from where candidate stars to be occulted by Pluto 
   and its satellites were determined, also based on data obtained with the WFI and 
   at dates close to those of the observations presented here. The analysis
   of the position of one of these candidates, as derived from a real
   occultation \citep{2011AJ....141...67S}, gives support to that value of 60 mas.
   This reinforces the fact that the levels of significance given by columns 
   15 and 16 in Table~\ref{table1} are not overestimating the accuracy of the
   observed ICRF sources' positions.

   \citet{2003AJ....125.2728A} report accurate optical positions of 172 ICRF 
   sources using a preliminary version of the UCAC2 as astrometric reference. We
   selected, from that paper, all those sources with a significant optical minus
   radio positional offset with existing structure information in the BVID. By 
   significant, we mean an offset whose absolute
   value in either right ascension or declination is greater than $3\sigma$ , where 
   $\sigma$ was defined in Section 4 making $\sigma_{D}$ equal to zero. 
   Table~\ref{table4} presents data about these sources. The structure indices to the 
   sources shown in Table~\ref{table4} are from the BVID and were selected by considering 
   that the dates of the VLBI experiments should be the closest to those of the respective optical 
   observations as given by \citet{2003AJ....125.2728A}.

\begin{table*}
\caption{Results from \citet{2003AJ....125.2728A}}             
\label{table4}      
\begin{center}          
\begin{tabular}{c c c c c c c c c c c c c c}     
\hline\hline       
        & \multicolumn{2}{c}{\hskip -6pt Struc. Ind.} & & & & & &
          \multicolumn{2}{c}{\hskip  3pt(Optical$-$VLBI)} & 
          \multicolumn{2}{c}{\hskip  3pt Internal} & 
          \multicolumn{2}{c}{\hskip  3pt 3$\sigma$ level} \\
 IERS ID   & X & S & Def.? & {\it gof} &   V  & Type   &    z    &
\hskip 8pt$\Delta\alpha{*}$ & \hskip 5pt$\Delta\delta$ & 
$\sigma_{\alpha}{*}$ & $\sigma_{\delta}$ &
$\sigma_{\alpha}{*}$ & $\sigma_{\delta}$ \\ 
\hline                 
0229$+$131 & 2 & 1 & Y & 8.9 & 17.7 & Q & 2.060 & $-$018 & $+$068 &  9 &  9 & 66 & 66 \\  
0827$+$243 & 2 & 1 & Y & 1.2 & 17.3 & Q & 0.940 & $-$015 & $-$085 &  9 &  4 & 66 & 61 \\
1252$+$119 & 2 & 1 & Y & 4.2 & 16.2 & Q & 0.873 & $-$036 & $-$072 &  7 &  6 & 64 & 63 \\
1532$+$016 & 4 & 2 & N & -   & 18.1 & Q & 1.435 & $+$054 & $-$068 &  9 &  4 & 66 & 61 \\
1546$+$027 & 3 & 1 & Y & 2.8 & 16.8 & Q & 0.414 & $+$024 & $-$075 & 11 &  4 & 68 & 61 \\
1548$+$056 & 2 & 2 & Y &55.5 & 17.7 & Q & 1.422 & $+$085 & $-$036 &  7 & 13 & 64 & 72 \\
1937$-$101 & 3 & 1 & N & 2.7 & 19.0 & Q & 3.787 & $-$017 & $-$067 & 12 &  4 & 70 & 61 \\
\hline                   
\end{tabular}
\end{center}          
Note: All angular values are in mas. The structure indices are from the BVID and were 
selected by considering that the dates of the VLBI experiments should be the closest to 
those of the respective optical observations ($\sim2000$ to the present table).
All columns are explained in Table~\ref{table1}. 
Columns from 9 to 12 were taken from table 4 given by \citet{2003AJ....125.2728A}.
\end{table*} 

   It is possible to see that 3 sources, out of the 7 shown in Table~\ref{table4}, have
   X-band structure indices of 3 or 4. 
   
   These mean epochs range from 1999 to 2001. 
   Source 1548$+$056 has a mean epoch of observation of 2000.20 and an X-band structure index of 
   2 in 1997 and of 3 in 2005. In this context, one could then argue that 4 (instead of 3) 
   sources, out of the 7 shown in Table~\ref{table4}, have X-band structure indices of 3 or 4.

   Therefore, as far as those sources (Tables~\ref{table1} and ~\ref{table4}) 
   with significant offsets are concerned, we can
   share the conclusions from \citet{2010IAU..POSTER.C8Z}\footnote{Poster presented 
   at the XXVII GA-IAU, Commission 8, where a short talk communicating the first 
   results of this work was also given.}: either a problem in the optical representation 
   of the radio frame (case one) or a real offset between the optical and radio positions of 
   the ICRF2 sources (case two).

\subsection{First case}

   As mentioned before, a value of 20 mas was adopted to the systematic errors in the UCAC2, 
   which is also in agreement with a previous estimate from an independent work 
   \citep{2003AJ....125.2728A}. If we trust in this value, the significant offsets would then 
   result mostly from systematic effects present in the Tycho-2 catalogue 
   \citep{2000A&A...355L..27H,2000A&A...357..367H}, the astrometric reference for the UCAC 
   series. 

   Systematic errors in positions and proper motions in the Tycho--2 were 
   studied with the help of common stars with the Hipparcos \citep{1997ESASP1200.....P} 
   in cells of $6^{\rm o}\times6^{\rm o}$. \citet{2000A&A...357..367H} show that these 
   systematic errors are below 1 mas and 0.5 mas/yr, for positions and proper motions 
   respectively, on angular scales of $6^{\rm o}\times6^{\rm o}$ or more. It is 
   interesting to mention here, however, that the 
   field of view for the UCAC observations was of about $1^{\rm o}\times1^{\rm o}$, much
   smaller than the areas used to check the Tycho--2 for systematic errors.

   It is through catalogues like the UCAC2 that most of the astronomical community 
   has access, in optical wavelengths, to the coordinate axes materialized by the ICRF. 
   If it contains systematic errors, no matter the origin, this is clearly of great 
   relevance. For instance, they could mimic a physical effect where it, in fact, does
   not exist.

\subsection{Second case}

   On the other hand, Fig.~\ref{figure6} shows that significant 
   offsets are most frequently associated with sources that have large (3 and 4) X-band
   structure indices. This is reinforced by the {\it gof} values presented in Table~\ref{table1}
   and would suggest a case of a real non-coincidence between the optical and 
   radio centers of the concerned sources. 

   In fact, optical and radio positions may differ when the emission is not core dominated
   \citep{2003Ap&SS.286..255D}. Most of the radio flux emission is associated to non-thermal 
   emission from jets, while the bulk of the optical emission is associated to the central 
   engine\footnote{See \citet{2009astro2010S.143J} for a brief enumeration of potential sources 
   of optical emission in quasars.}. This must imply on parsec scale separation between the 
   two emitting regions. Adopting the Standard Model, this translates to a 
   floor of sub mas angular distances \citep{2000A&A...359.1201F}. Accordingly, 
   \citet{2008A&A...483..759K} estimate an average shift between the radio and optical 
   cores of 0.1 mas from theoretical assumptions on the optical emission. This value is two 
   to three orders of magnitude smaller than the significant offsets shown in 
   Table~\ref{table1} and Table~\ref{table4}. In this context, our results (and those of
   Table~\ref{table4}, to some extent) are a 
   relevant material to better understand the origin of optical emission in quasars. 

   The investigation of a larger number of sources would, undoubtfully, help to decide
   which case most contributes to the significant offsets found here.

\subsection{Considerations about the time dependence of the structure index}

   It is just as well to mention that the structure indices of the sources may change as
   a function of time and that not all sources were observed close to the dates when these
   indices were determined (see Tables~\ref{table2} and~\ref{table3}). Table~\ref{table3}
   reports, as obtained from the BVID, data from the 3 most recent years of experiments
   (when available). It is possible to notice that the time variation 
   of these indices, as far as the sources here studied are concerned, typically do not exceed 
   one. This is
   often true also when we consider, when available to the same experiments, dates older 
   than those from the 3 most recent years in the BVID. One clear 
   exception is the source 0440$-$003. A. Collioud (2011, personnal comm.) determined the 
   continuous X-band structure index \citep{2011A&A...526A.102B} to this source and found
   the value of 1.83 to the experiment of 1995. This indicates, therefore, a variation
   from about 2 to 4 which is less pronounced than that of 1 to 4. In general, it is then 
   reasonable to assume that the results pointed out by Fig.~\ref{figure6} remain the 
   same although observation dates are not always close to those of the respective VLBI 
   experiments.

\subsection{Comments on individual sources}

   Source 0743$-$673 has its position given by both SOAR and ESO/MPG 
   telescopes, where a difference of 43 mas and 30 mas is seen, respectively,
   between the right ascension and declination as determined from both 
   instruments. It should also be noticed that the number of
   reference stars used to obtain the final position of this source from
   the ESO/MPG is 380, whereas only 8 were used to obtain the final position
   from the SOAR. It is just as well to remember that the area on the sky
   covered by the WFI is about 30 times that of the SOI. In this context,
   a possible important contribution for the above difference comes from the
   materialization of the celestial frame as given by two different sets of
   reference stars, the smallest one giving rise to statistical fluctuations. 

   Source 0754$+$100 was also observed by both telescopes. In this case, one
   can find differences of 46 mas and 22 mas, respectively, between the right 
   ascension and declination as determined from both 
   instruments. The number of reference stars used to obtain the source's
   position from the WFI (408) and that to obtain the source's position
   from the SOI (9) leads to the same rationale presented to 0743$-$673
   to explain the differences. It is also noticed that, from the observations
   of 0754$+$100 made with SOI only, this source should appear in 
   Fig.~\ref{figure6} as a circle. The results from the WFI, however, come
   from a much larger number of reference stars spread on a larger area of the
   sky. Therefore, the offsets and respective significance levels obtained
   from this last imager were preferred.  

   To both sources (0743$-$673 and 0754$+$100), the values of $\Delta\alpha{*}$ 
   and $\Delta\delta$ as obtained from the positions given by the SOAR and the 
   ESO/MPG telescopes are within their respective uncertainties at the $1\sigma$ 
   level.

   A well known strategy in astrometry, aiming at avoiding systematic errors in the
   determination of positions of celestial bodies, is to observe (and link) objects 
   separated by large angles on the sky. This is not, by far, the case for any of 
   the two imagers considered in this work. It is reasonable, at least, to expect 
   that the larger the number of reference stars on the image the better the celestial
   frame can be represented. A large number of reference stars was always available
   on the WFI frames which is not verified in the case of the SOI. This fact was
   not taken into consideration when determining the significant thresholds 
   (last two columns of Table~\ref{table1}) to the offsets of each source. 
   The offsets obtained to sources 0743$-$673 and 0754$+$100, as given by the SOI and the WFI, 
   give a rough indication of how different results can be. In any case, as mentioned above,
   these differences are within the uncertainties at the $1\sigma$ level.

\section{Conclusion}

   We have determined accurate positions for 22 ICRF2 sources in optical
   wavelengths with the SOAR and the ESO/MPG telescopes. A conservative level
   of significance to the optical to radio offsets in position was
   determined to each source, as given by Table~\ref{table1}. Four sources
   presented significant offsets.

   We understand, as also suggested by \citet{2010IAU..POSTER.C8Z}, that these 
   significant offsets indicate either a problem with the optical representation 
   of the ICRF or a real offset between the respective optical and radio positions. 

   In the first case, the UCAC2 is directly addressed. Assuming that this catalogue
   represents the IAU's celestial frame as materialized by the Tycho--2
   catalogue within the claimed accuracy \citep[see][]{2004AJ....127.3043Z,
   2010AJ....139.2184Z}, then it is the Tycho--2 that presents systematic errors. 

   In the second case, we can suppose that the results outlined by Fig.~\ref{figure6}
   are not just a coincidence and that significant optical
   to radio offsets in position are related to high (3 and 4) X-band structure 
   indices. In this context, the results presented here would be of interest 
   in studying the origin of optical emission in quasars.

   Whatever the case, the consequences are of importance. It is through catalogues
   like the UCAC2 that most of the astronomical community has access, in optical
   wavelengths, to the coordinate axes materialized by the ICRF. The existence
   of systematic errors may lead, for instance, to erroneous conclusions about the physics 
   of a given object.
   On the other hand, assuming that the significant offsets found here are 
   definitely correlated to the X-band structure index then the
   relation between optical and radio frames in the future must consider
   spatially extended structure effects.

   A larger number of accurate
   optical positions of reference frame sources would greatly help to
   tell from which case the results presented here come from.

\begin{acknowledgements}
J.I.B.C., A.H.A., M.A. and R.V.M. acknowledge CNPq grants 151392/2005-6, 
477943/2007-1, 478318/2007-3, 306028/2005-0, 304124/2007-9 and 307126/2006-4. 
J.I.B.C., M.A., and D.N.S.N. thank FAPERJ for grants E-26/170.686/2004, 
E-26/100.229/2008 and E-26/110.177/2009. A.H.A. thanks Marie Curie fellowship 
grant PIIF-GA-2009-236735. This publication makes use of data products from 
the Two Micron All Sky Survey, which is a joint project of the University of 
Massachusetts and the Infrared Processing and Analysis Center/California Institute 
of Technology, funded by the National Aeronautics and Space Administration and the 
National Science Foundation. This research has made use of the VizieR catalogue 
access tool, CDS, Strasbourg, France. This research has made use of NASA's 
Astrophysics Data System Bibliographic Services.
The authors acknowledge an anonymous referee for fruitful discussions and
suggestions.
\end{acknowledgements}

\bibliographystyle{aa}
\bibliography{camargo_revised_2col}

\end{document}